# A nitrogen-vacancy spin based molecular structure microscope using multiplexed projection reconstruction.

Andrii Lazariev[1] and Gopalakrishnan Balasubramanian[1,2] *.

[1] MPRG Nanoscale Spin Imaging, Max Planck Institute for Biophysical Chemistry, Göttingen, Germany.

[2] Center Nanoscale Microscopy and Molecular Physiology of the Brain (CNMPB), Göttingen, Germany.

*Correspondence and requests for materials should be addressed to G.Balasubramanian (gbalasu@mpibpc.mpg.de)

**ABSTRACT**

**Methods and techniques to measure and image beyond the state-of-the-art have always been influential in propelling basic science and technology. Because current technologies are venturing into nanoscopic and molecular-scale fabrication, atomic-scale measurement techniques are inevitable. One such emerging sensing method uses the spins associated with nitrogen-vacancy (NV) defects in diamond. The uniqueness of this NV sensor is its atomic size and ability to perform precision sensing under ambient conditions conveniently using light and microwaves (MW). These advantages have unique applications in nanoscale sensing and imaging of magnetic fields from nuclear spins in single biomolecules. During the last few years, several encouraging results have emerged towards the realization of an NV spin-based molecular structure microscope. Here, we present a projection-reconstruction method that retrieves the three-dimensional structure of a single molecule from the nuclear spin noise signatures. We validate this method using numerical simulations and reconstruct the structure of a molecular phantom β-cyclodextrin, revealing the characteristic toroidal shape.**

## Introduction

Nuclear magnetic resonance (NMR) is a widely applied technique that infers chemical signatures through magnetic dipolar interactions. Magnetic fields arising from nuclear spins are weak, so conventional NMR measurement requires a sizable number of spins, approximately $10^{15}$, to achieve reasonable signal-to-noise ratio (SNR)[1]. These sensitivity limitation permits only ensemble-averaged measurements and forbids any possibilities of studying individual molecules or their interactions. There is a continuous effort to use hybrid detection strategies to improve the sensitivity and, thus, achieve single molecular sensing with spin information. Spin-polarized scanning tunneling microscopy (STM)[2] and magnetic resonance force microscopy (MRFM)[3] have displayed remarkable abilities in imaging structures with chemical contrast to single electron spins[3] and few thousand nuclear spins[4]. Their extreme sensitivity imposes restrictions on the permissible noise floor, so the microscope is operable under cryogenic conditions.

It was the first room-temperature manipulation of a single spin associated with the nitrogen-vacancy (NV) defects in diamond[5] that brought the NV center into quantum limelight. In this seminal paper, Gruber *et al.* envisaged that material properties can be probed at a local level using optically detected magnetic resonance (ODMR) of NV spin combined with high magnetic field gradients ensemble averaging[5]. Later, Chernobrod and Berman proposed scanning-probe schemes to image isolated electron spins based on ODMR of photoluminescent nanoprobes[6]. Perceiving the uniqueness of single NV spin and combining coherent manipulation schemes[7,8], independently proposals[9,10] and experimental results[11,12] emerged, ascertaining NV spin as an attractive sensor for precision magnetometry in nanoscale[13–15].

Spin sensor based on NV defects is unique because it is operable under ambient conditions and achieves sufficient sensitivity to detect few nuclear spins[16–20]. The hydrogen atoms that are substantially present in biomolecules possess nuclear spins. Mapping spin densities with molecular-scale resolution would aid in unraveling the structure of an isolated biomolecule[21,22]. This application motivates to develop an NV spin-based molecular structure microscope[23–26]. The microscope would have immense use in studying structural details of heterogeneous single molecules and complexes when other structural biology tools are prohibitively difficult to use. For example, NV spin-based molecular structure microscope would find profound implications in the structure elucidation of intrinsically disordered structures, such as the prion class of proteins (PrP). This family of proteins is known to play a central role in many neurodegenerative diseases[27]. Understanding the structure-function relationship of this protein family will be crucial in developing drugs to prevent and cure these maladies.

The NV spin is a high dynamic range precision sensor[28,29]; its bandwidth is limited only by the coherence time and MW driving-fields[30,31]. The broadband sensitivity has an additional advantage because multiplexed signals can be sensed[32–34]. Fully exploiting this advantage, we present a projection-reconstruction method pertaining to an NV spin microscope that encodes the spin information of a single molecule and retrieves its three-dimensional structure. We analyze this method using numerical simulations on a phantom molecule β-cyclodextrin. The results show distinct structural features that clearly indicate the applicability of this technique to image isolated biomolecules with chemical specificity. The parameters chosen for the analysis are experimentally viable[4,35–37], and the method is realizable using state-of-the-art NV sensing systems[23,24,35]. We also outline some possible improvements in the microscope scheme to make the spin imaging more efficient and versatile.

At equilibrium, an ensemble of nuclear spins following the Boltzmann distribution tends to have a tiny fraction of spins down in excess of spins up. The expression for this population difference is given by the Boltzmann equation:

$$\Delta N = N\left(e^{\Delta E/kT} - 1\right) \approx N \frac{h\gamma B}{2\pi kT}, \qquad (1)$$

where *N* is the number of spins, *ΔE* is the energy level difference, *k* is the Boltzmann constant, *T* is the temperature, *h* is the Planck constant, *γ* is the gyromagnetic ratio, and *B* is the magnetic field. For room temperature and low-field conditions, ($\Delta E \ll kT$) an approximation is made in equation (1). The spins reorient their states (↑-↓, ↓-↑) with a characteristic time constant while conserving this excess population. The average value of this excess spins remains constant while the root mean square (r.m.s.) value of this fluctuation over any period is given by $\sigma \propto \sqrt{N}$. This statistical fluctuation in the net magnetization signal arising from uncompensated spins is called spin noise, and it is relevant when dealing with small numbers of spins[38]. Combining spatial encoding using the magnetic field gradients and passively acquiring spin noise signals, Müller and Jerschow demonstrated nuclear specific imaging without using RF radiation[39]. The spin noise signals become prominent when we consider fewer than $10^6$ spins. Considering solid-state organic samples with spin densities of $5\times10^{22}$ spins/cm$^3$, this quantity of spins creates a nanoscale voxel[36]. Spin noise signals arising from a few thousand nuclear spins were sensed using MRFM and used to reveal a 3D assembly of a virus[4]. More recently, using NV defects close to the surface of a diamond, several groups were able to detect the nuclear spin noise from molecules placed on the surface under ambient conditions[19,20,25].

If a magnetic field sensor is ultra-sensitive and especially sample of interest is in nanoscale[13], it is convenient to use spin noise based imaging. The advantage being, this sensing method does not require polarization or driving nuclear spins. Dynamic decoupling sequences, such as (XY8)$_{n=16}$, are used for sensing spin noise using NV. The method uses pulse timings to remove all asynchronous interactions and selectively tune only to the desired nuclear spin Larmor frequency[17,19,25]. The NV coherence signal is recorded by varying the interpulse timings over a desired range. The acquired signal is deconvoluted with the corresponding filter function of the pulse sequence to retrieve the power spectral density of noise or the spin noise spectrum[40]. This approach has been proved to be sensitive down to a single nuclear spin in the vicinity[20] as well as to a few thousand of nuclear spins at distances exceeding 5 nm [19]. The spin sensing results clearly demonstrate the potential of NV spin sensor as a prominent choice for realizing a molecular structure microscope that is operable under ambient conditions.

**Spin imaging method**

Several different schemes are currently being considered for nanoscale magnetic resonance imaging (MRI) using NV-sensors: scanning the probe[12,15,41], scanning the sample[23,24] and scanning the gradient[12,35]. The first two rely on varying the relative distance and orientation of the sensor to the sample, thereby sensing/imaging the near-field magnetic interaction between the NV spin and the spins from the sample. This sensing could be performed either by passively monitoring the spin fluctuation[19] or by driving sub-selected nuclear spins by RF irradiation[18]. This approach is particularly suitable when dilute spins need to be imaged in a sample or for samples that are sizable. These methods read spin signal voxel-by-voxel in a raster scanning method, so they are relatively slow but have the advantage of not needing elaborate reconstruction[23,24]. For the method presented here to image a single molecule, we employ scanning the gradient scheme[12,35].

Here, we present a three-dimensional imaging method that is especially suiting for nanoscale-MRI using NV spins. The multiplexed spin-microscopy method uses a projection-reconstruction technique to retrieve the structure of a biomolecule. Fig. 1a shows a schematic of the setup that is similar to those utilized in nanoscale magnetometry schemes[12,35,42]. We place the sample of interest (a biomolecule) on the diamond surface very close to a shallowly created NV defect[43]. Achieving this could be perceived as a difficult task, but recent advancements in dip pen nanolithography (DPN)[44] and micro-contact printing (μCP)[45] for biomolecules have been able to deposit molecules with nanometer precision.

The encoding stage proceeds in the following manner: a shaped magnetic tip is positioned such that we subject the sample to a magnetic field gradient on molecular scales (2 – 5 G/nm)[4,35]. In this condition, nuclear spins present in the biomolecule precess at their Larmor frequencies depending on their apparent positions along the gradient (Fig. 1b). The spin noise spectrum from the sample is recorded using an NV center in close proximity[19]. Because of the presence of a field gradient on the sample, fine-features appear in the noise spectrum. These unique spectral signatures correspond to nuclear spin signal contributions from various isomagnetic field slices[39] (Fig. 1c,d). Directions of field gradient applied to the molecule are changed by moving the magnetic tip to several locations. This gradient gives distinct projection perspectives while the spin noise spectrum contains the corresponding nuclear spin distribution information. These signals are indexed using the coordinates of the magnetic tip with respect to the NV defect as $\Theta$ and $\Phi$ (measured using high--resolution ODMR) and are stored in a 3D array of $S(\Theta,\Phi,\omega)$ values. We require encoding only in one hemisphere because of the linear dependence (thus redundancy) of the opposite gradient directions. For a simple treatment, we assume the gradients to have small curvatures when the magnetic tip to NV sensor distances are approximately 100 nm (i.e. a far field). This assumption is along the lines of published works and is valid for imaging biomolecules of approximately 5 nm in size at one time[4,35].

The reconstruction procedure is as follows: by knowing the complete magnetic-field distribution from the tip[42], we can calculate the gradient orientation ($\theta, \varphi$) at the sample location for any tip position ($\Theta, \Phi$) and rescale the spectral information to spatial information in 1D: $\omega = \gamma r \nabla B$. In this way, the encoded data set matrix dimensions are transformed into $\theta, \varphi$ and $r$.

$$s(r, \theta, \varphi) \propto \iiint B_{rms}(x, y, z)\, \delta(x\sin\theta\cos\varphi + y\sin\theta\sin\varphi + z\cos\theta - r)\, dxdydz \quad (2)$$

Here, $B_{rms}(x, y, z)$ is the magnetic field fluctuation caused by the number of nuclear spins $N(x, y, z)$ contained in the respective isomagnetic field slices. As nuclear spin Larmor frequencies within every slice are identical, they all contribute to same frequency component in noise spectrum[39,46].

Therefore, the 1D signal we recorded is an integrated effect from the spins in the respective planes (refer to equation (2)). The next procedure follows along the lines of a filtered back-projection principle to remove high-frequency noise and projection artifacts. Here, we should ensure appropriate quadratic filters because the signal is a plane integral but not a line integral as in X-ray computed tomography (CT)[46]. The rescaled signal $s(r, \theta, \varphi)$ is filtered using a quadratic cutoff in the frequency domain. We perform the reconstruction in the following way; an image array is created with $n^3$ dummy elements in three dimensions $I(x, y, z)$. Any desired index $r_i, \theta_j, \varphi_k$ from the signal array is chosen, and the corresponding value $s(r_i, \theta_j, \varphi_k)$ is copied to the image array at location index $z = r_i$. The values are normalized to $n$ and replicated to every cell in the xy-plane at the location $z = r_i$. The elements of the xy-plane are rotated to the values $-\theta_j, -\varphi_k$ following the transformation given by affine matrices shown in equation (3).

$$\begin{bmatrix} x' \\ y' \\ z' \\ 1 \end{bmatrix} = \begin{bmatrix} \cos(-\theta_j) & -\sin(-\theta_j) & 0 & 0 \\ \sin(-\theta_j) & \cos(\theta_j) & 0 & 0 \\ 0 & 0 & 1 & 0 \\ 0 & 0 & 0 & 1 \end{bmatrix} \begin{bmatrix} \cos(-\varphi_k) & 0 & \sin(-\varphi_k) & 0 \\ 0 & 1 & 0 & 0 \\ -\sin(-\varphi_k) & 0 & \cos(\varphi_k) & 0 \\ 0 & 0 & 0 & 1 \end{bmatrix} \begin{bmatrix} x \\ y \\ z \\ 1 \end{bmatrix} \quad (3)$$

The values are then cumulatively added to the dummy elements and stored in the transformed index. This procedure is repeated for every element in the signal array so that the corresponding transformed array accumulates values from all of the encoded projections. The transformed array with units in nm contains raw projection-reconstructed images. This array is then rescaled to account for the point spread function of NV spin and single proton interaction[23]. The resulting 3D matrix carries nuclear spin density in every element and contains three-dimensional image of the molecule.

**Results**

To evaluate this technique, we considered a simple molecule of β-cyclodextrin as a molecular phantom. This molecule has a toroidal structure with an outer diameter of 1.5 nm and an inner void of 0.6 nm (Fig. 2a). We specifically selected this molecule so that we could easily visualize the extent of the structural details that can be revealed by reconstruction. We used the crystallographic data of β-cyclodextrin from the Protein Data Bank and considered the coordinate location of all the hydrogen atoms for the numerical simulations performed using MATLAB. Some relevant information about the molecular spin system and parameters used for the simulations is listed in Table 1.

We virtually position the molecule in the proximity of an NV defect that is placed 5 nm beneath the surface of the diamond. At these close distances, the spin noise from hydrogen atoms is sensed by the NV spin[19]. We compute the fluctuating magnetic field amplitude (r.m.s.) produced by proton spins at the location of the NV spin by using the expression given by Rugar et al[23]. The β-cyclodextrin molecule, when placed in the vicinity, produces a field of about 94 nT (r.m.s.), matching reported values[19]. We subject the molecule to the magnetic field gradients of 3 G/nm, and this produces spread in Larmor frequencies of 30.6 kHz for the hydrogen spins. As explained before, we compute the $B_{rms}$ field produced by hydrogen spins in each isofield slice set by the spectral resolution ~ 1.3 kHz. The $^1$H spins in respective slices precess at their characteristic Larmor frequencies, so the noise spectrum reveals spectral features containing information about spin density distribution along the gradient direction. We show the computed spin noise spectra ($B_{rms}$ vs. frequency shifts) from the β-cyclodextrin molecule representing two different gradient orientations in Fig 2b.

For three-dimensional encoding and reconstruction, we considered 9x9 unique gradient orientations equispaced along different $\theta, \varphi$ angles. The spectral data are computed for every projection, converted to spatial units and stored in a 3D matrix. This signal matrix is shown in Fig. 2c as slices in the $r, \theta$ dimensions. We apply the reconstruction algorithm as explained above to get the spatial distribution of hydrogen atoms. The structure of the reconstructed molecule clearly reveals its characteristic toroidal shape (Fig. 2d). The quality of the image reconstruction depends on the number of distinct tip locations (or gradient orientations) used for encoding[39]. The simulations presented in Fig. 2 display the reconstruction quality achieved for a toroidal molecule, β-cyclodextrin, for a set of 81 measurements used for encoding and decoding. If we consider the signal acquisition time of 22 seconds reported for a single point spectral measurement[23] and calculate the time needed for achieving desired spectral resolution (~1.3 kHz), it results in long averaging times. We note that the signal acquisition time dramatically reduces to ~ 1 second/point by using double-quantum magnetometry[47] together with enhanced fluorescence collection[48]. In this case the complete image acquisition time becomes approximately 33 minutes for the data set used here to reconstruct the molecular structure of an isolated β-cyclodextrin.

**Discussion**

The β-cyclodextrin molecule we have considered for simulations is a simple molecule, but it has a characteristic toroidal structure and is easy to visualize. The results clearly showed molecular-scale resolution and provided information about the structure. The structural details and achievable resolution depends on the following factors: The primary factor is the SNR obtainable when recording the noise spectra. Demonstrations using double quantum transitions ($m_s$=-1 ⇔ $m_s$=+1) achieved high-fidelity spin manipulation[49] and improved sensing[47]; applying those techniques could improve signal quality. Spectral reconstruction using compressed sensing approach is expected to considerably speed-up sensing[21,22,33,34]. In addition to this, improved fluorescence collection efficiency from single NV defects can be achieved using nanofabricated pillars[50], solid immersion lenses[51] and patterned gratings[48]. The primary source of noise being the photon shot noise, boosting signal quality naturally increases the achievable resolution. Other techniques, such as dynamic nuclear polarization[52], selective polarization transfer[17,19,20,22], the quantum spin amplification mediated by a single external spin[53], could give additional signal enhancements. Schemes employing ferromagnetic resonances to increase the range/sensitivity could provide other factors for resolution improvement[54,55]. The presented method is efficient for the reconstructing structural information and spins distribution at the nanoscale whenever it is possible to perform three-dimensional encoding. The encoding can be done either by using an external gradient source[12,35,56] or by the field gradient created by NV in its vicinity[20,57].

It is important to consider nuclear spins from water and other contaminants that form adherent monolayers on the surface of diamond[4,19,23,24]. The gradient encoding will register their spatial location appropriately. Upon reconstruction, this would result in a two-dimensional layer seemingly supporting the molecule of interest. This plane could come as a guide for visualization but can be removed by image processing if required. Homo-nuclear spin interactions cause line broadening and become a crucial factor when dealing with spins adsorbed on a surface. Unwanted spin interactions can be minimized by applying broadband, and robust decoupling methods such as phase modulated Lee-Goldburg (PMLG) sequences[58].

The factors determining the achievable structural resolution are the magnetic field gradient, the SNR of spin signals, number of distinct perspectives and the spectral linewidth of the sample. It is practical to retrieve structural details with atomic resolutions by applying larger gradients, using SNR enhancing schemes, improving fluorescence collection, and using decoupling. We have considered reported parameters and demonstrated that our method can achieve molecular-scale resolution. Although continued progress clearly indicates that attaining atomic resolutions is within reach[35,59]. However, for many practical applications, it is sufficient to obtain molecular-scale structures that contain information relevant to biological processes. The key feature of the NV-based molecular structure microscope is the ability to retrieve the three-dimensional structural details of single isolated biomolecules under ambient conditions without restrictions on the sample quality or quantity. These will be very much useful for studying hard-to-crystallize proteins and intrinsically disordered proteins. A molecular structure microscope that has the potential to image molecules like prion proteins would be pivotal in understanding the structure, folding intermediates and ligand interactions. These insights would undoubtedly pave ways to understand the molecular mechanisms of diseases pathways and develop efficient therapeutic strategies for their treatment and prevention[60].

**Acknowledgements**

We gratefully acknowledge funding from the Max-Planck Society, Niedersächsisches Ministerium für Wissenschaft und Kultur and DFG Research Center Nanoscale Microscopy and Molecular Physiology of the Brain.

**Author contributions**

A.L and G.B conceived the method. A.L performed the numerical simulations and both authors analysed the results. G.B wrote the paper with input from A.L. Both authors reviewed the manuscript.

**Competing financial interests**

The authors declare no competing financial interests.

**Correspondence and requests for materials should be addressed to**

G.Balasubramanian (gbalasu@mpibpc.mpg.de)

**Figure 1 | Schematics of the molecular scale spatial encoding using magnetic field gradients.** (a) Schematic representation of a biomolecule in the vicinity of an NV-center; $\omega_L$ in the inset signifies the Larmor peak position in the spectrum. (b) Gradient encoding of the molecule's proton density; a spectrum is presented in direction of the gradient vector. (c),(d) Schematic representation of different magnetic field gradients induced by an approached magnetic tip; insets: their influence on the spectrum.

**Figure 2 | Simulations of projection-reconstruction method using a molecular phantom β-cyclodextrin.** (a) 3D visualization of hydrogen atoms in β-cyclodextrin molecule in a space filling representation. (b) Simulated spin noise spectra for two different gradient orientations. (c) Encoded signal matrix (some slices are omitted for visual clarity). (d) Reconstructed three dimensional image of a β-cyclodextrin molecule.

**Table 1 | Relevant parameters for imaging a molecular phantom β-cyclodextrin using NV spin based molecular structure microscope by projection-reconstruction method.**

| Quantity | | Value | |
| --- | --- | --- | --- |
| **Phantom molecule** | Molecule name | β-Cyclodextrin (Cycloheptaamylose) | |
| | Chemical formula | $C_{42}H_{70}O_{35}$ | |
| | Molecular weight | 1135 g/mol (1.14 kDa) | |
| | Molecular size | OD-15Å and ID- 6Å | |
| | Molecular shape | Toroidal | |
| | Proton density | $5.8 \times 10^{28}$ m$^{-3}$ (58 protons/nm$^3$) | |
| **B$_0$ Field** | B$_0$ Field | 500 Gauss | |
| | $^1$H Larmor | 2128.5 kHz | |
| **Encoding** | Gradient (magnetic tip) | 3 G/nm @ 100nm | |
| | Larmor frequency spread | 30.6 kHz | |
| | Spectral resolution ($\Delta f$) | 1.28 kHz | |
| **Signal B (r.m.s.)** | $10^4$ - $^1$H @ 7nm [19] | ~ 400nT | |
| | 70 - $^1$H @ 5nm | ~ 94nT | |
| **Signal acquisition time/point** | | *Δf = 1.3 kHz* | *Δf = 30 kHz* [23] |
| | $^1$H using (XY8)n [23] | 586 sec | 22 sec |
| | + DQC(XY8)n [47] | 37 sec | 1.3 sec |
| | + Enhanced collection [48,50] | 1 sec | 0.036 sec |
| **Projections** | Distinct projections | $\theta$ (0°-180°) and $\varphi$ (0°- 180°) | |
| | Total | 81 [9x9] | |
| **3D structure acquisition time** | | ~ 33 minutes | |

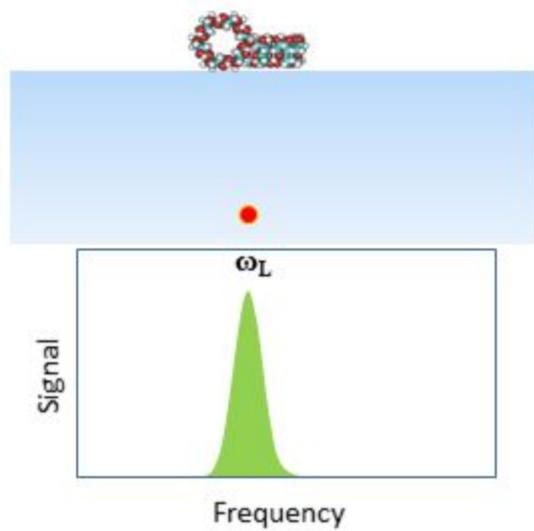
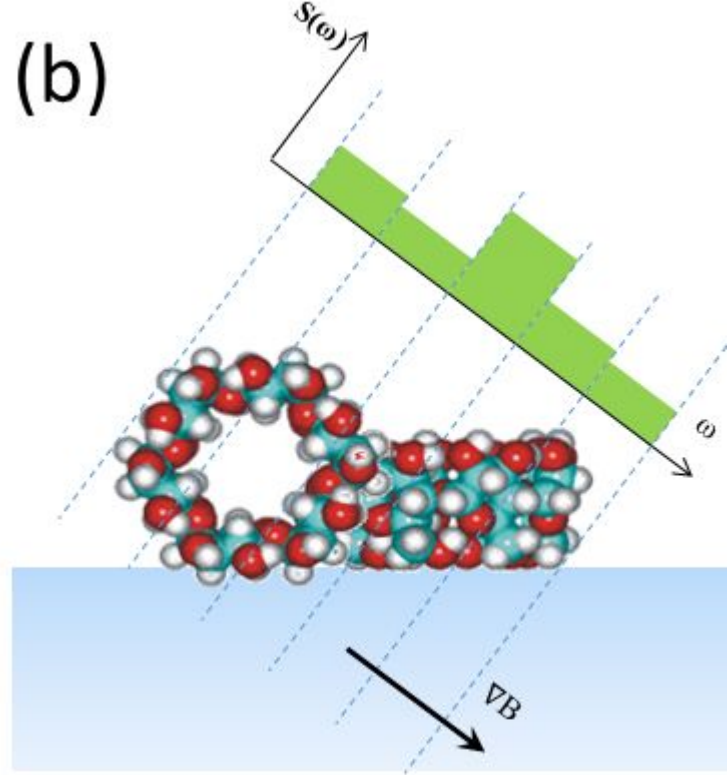
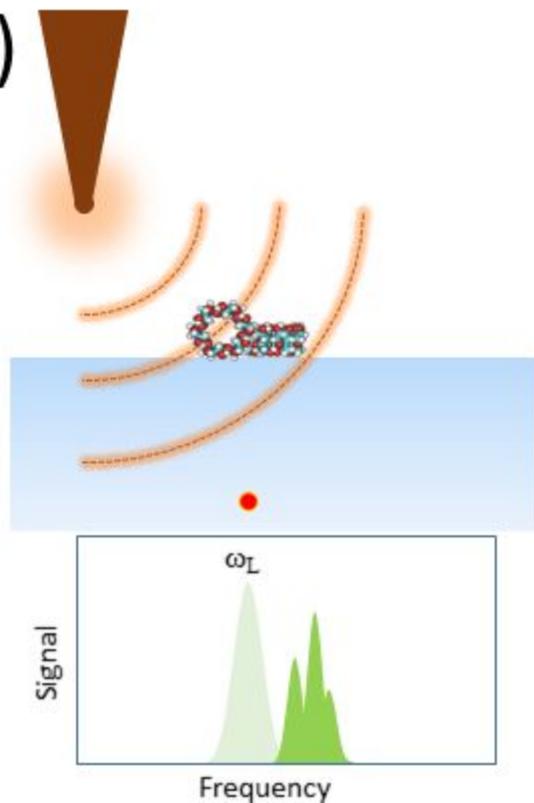
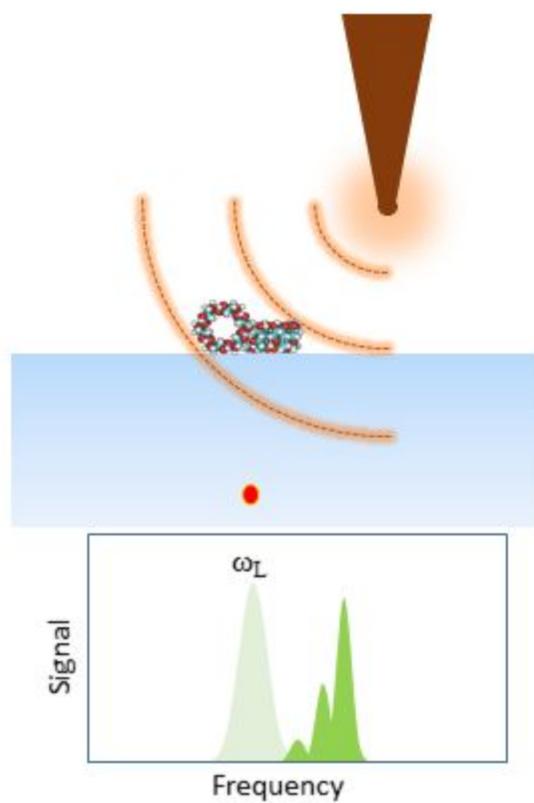

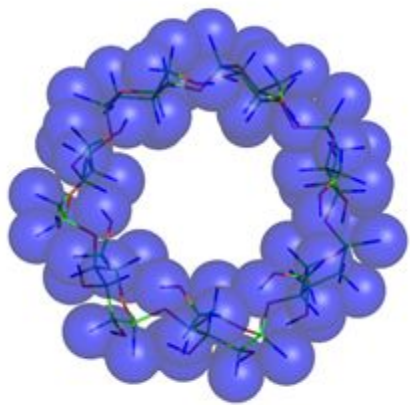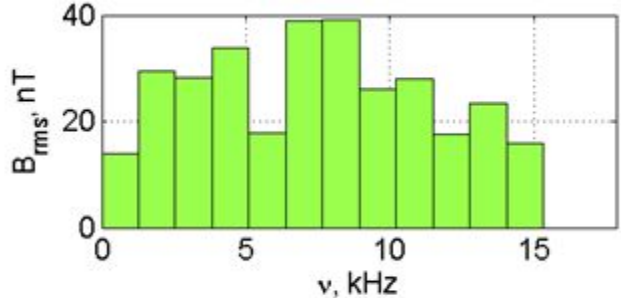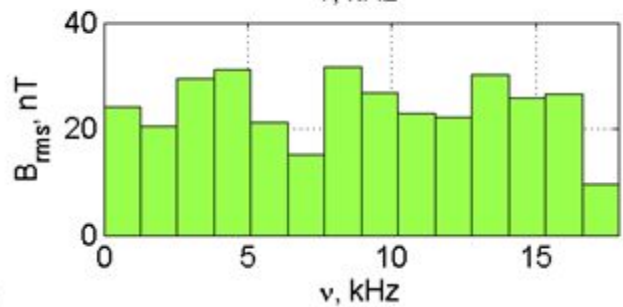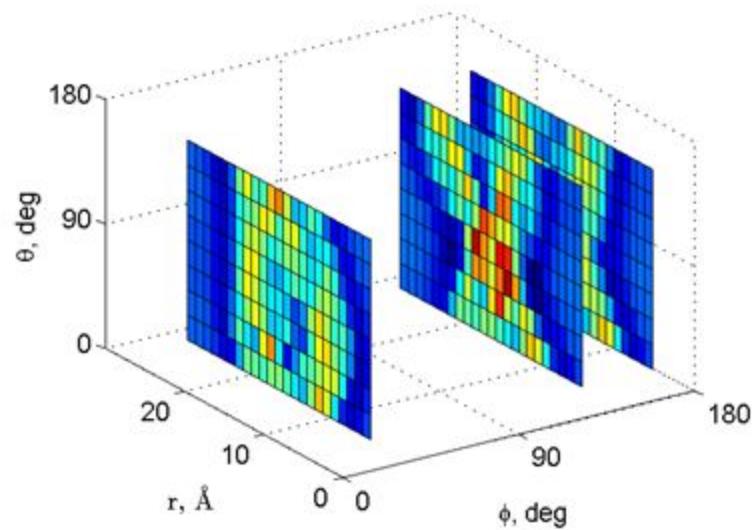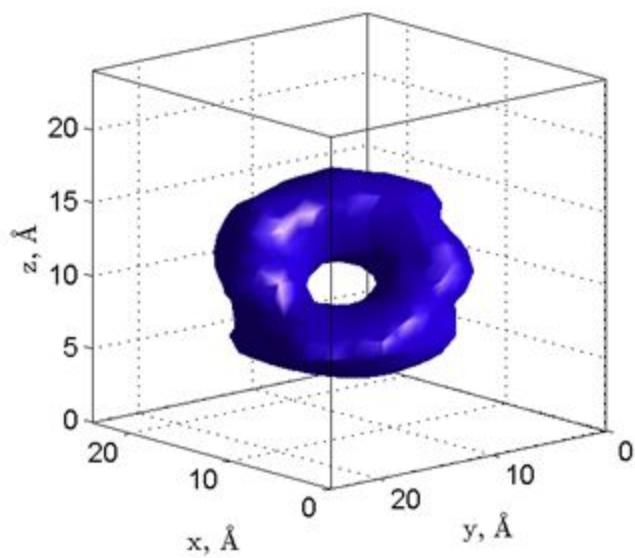